\documentclass[journal,twoside,web]{ieeecolor}
\usepackage{generic}
\usepackage{cite}
\usepackage{amsmath,amssymb,amsfonts}
\usepackage{graphicx}
\usepackage[linesnumbered,ruled,vlined]{algorithm2e}
\usepackage[hidelinks]{hyperref}
\usepackage{textcomp}
\let\labelindent\relax
\usepackage{enumerate}
\usepackage{caption}
\usepackage{subcaption}
\usepackage{subfiles}
\usepackage{xspace}
\usepackage{mathtools}
\usepackage[shortlabels]{enumitem}
\mathtoolsset{showonlyrefs}
\usepackage{svg}
\usepackage{slashbox}
\usepackage{xfrac}  
\usepackage[english]{babel}
\usepackage{soul}

\newcommand{\R}{\ensuremath{\mathbb{R}}\xspace}
\newcommand{\K}{\ensuremath{\mathcal{K}}\xspace}

\newcommand{\X}{\ensuremath{\mathcal{X}}\xspace}
\newcommand{\U}{\ensuremath{\mathcal{U}}\xspace}
\newcommand{\C}{\ensuremath{\mathcal{C}}\xspace}
\newcommand{\sys}{\ensuremath{\text{sys}}\xspace}

\newtheorem{theorem}{Theorem}[section]
\newtheorem{lemma}{Lemma}[section]
\newtheorem{problem}{Problem}

\newtheorem{definition}{Definition}[section]
\newtheorem{assumption}{Assumption}[section]

\newtheorem{corollary}{Corollary}[section]

\def\BibTeX{{\rm B\kern-.05em{\sc i\kern-.025em b}\kern-.08em
    T\kern-.1667em\lower.7ex\hbox{E}\kern-.125emX}}
\DeclareMathOperator*{\argmin}{arg\,min}

\begin{document}
\title{Verification of High-Order Control Barrier Functions by Computing Class $\K$ Functions}
\author{Ellie Pond and Matthew T. Hale$^1$
\thanks{$^1$ Authors are with the School of Electrical and Computer
Engineering, Georgia Institute of Technology, Atlanta, GA, USA.
Emails: \texttt{\{epond3,mhale30\}@gatech.edu}
}
\thanks{
This work was supported by 
ONR under grant N00014-22-1-2435,
AFOSR under grant FA9550-19-1-0169, 
and
AFRL under grant FA8651-22-F-1052. 
}
}

\maketitle

\begin{abstract}
High-order control barrier functions (HOCBFs) can 
enforce system safety, but 
it must be verified that a system can actually implement
a given HOCBF (or collection thereof). 
We address this need by formulating 
a sequence of SOS programs that verify HOCBFs by computing the class $\K$ functions
associated with them. We show that if solutions to these SOS programs exist,
then a system is guaranteed to remain safe at runtime. 
Simulations show this approach in practice.
\end{abstract}


\section{Introduction}
\label{sec:introduction}


Recently, control barrier functions (CBFs) have successfully been used to achieve safe operation of safety-critical systems \cite{ames2016control,Jankovic_CollisionAvoidance,nguyen2016exponential,xu2016control,garg2024advances}.
High-order control barrier functions (HOCBFs) can encode safety requirements,
and they generalize ordinary CBFs because they can
have non-trivial relative degree. 
HOCBFs 
require that a set of safe states is forward-invariant under a system's dynamics. 
Beyond a single HOCBF, other system specifications often must be accounted for in many settings.
These can include control Lyapunov functions (CLFs) that enforce goal satisfaction \cite{romdlony2014uniting}, input constraints that encode the physical limitations of a system \cite{agrawal2021safe}, and additional HOCBFs that encode separate safety requirements \cite{xu2016control}. 

At runtime, HOCBFs are often implemented by 
solving a sequence of optimization problems to produce safe inputs to a system,
and these problems incorporate a constraint for safety. 
This constraint contains a higher-order derivative of an HOCBF itself and several class $\K$ functions. 
These class $\K$ functions are user-determined and 
there is often a 
lack of rules
for choosing these functions 
to provide forward invariance~\cite{de2023synthesis,xiao2019control}.
Moreover, when computing inputs, it is common to assume that each
optimization problem is feasible, but it is not clear
when all of a system's specifications can be satisfied simultaneously \cite{xu2016control,romdlony2014uniting,garg2024advances}.
This need has led to the problem of CBF verification \cite{zhang2023exact,prajna2007framework,srinivasan2020synthesis,xiao2021HOCBFs,ma2022learning,xu2017correctness},
which ensures that safe inputs can be generated online at all times.

We present new techniques that verify HOCBFs by automatically constructing
the class~$\K$ functions that they require. 
We utilize the connection between inequality constraint satisfaction in optimization problems 
and the notion of nonnegativity certificates from
real algebraic geometry \cite[Chap. 5]{powers2021certificates}, which can
be computed efficiently with sums-of-squares programs (SOSPs) \cite{lasserre2015introduction}.
We construct SOSPs for various system scenarios to be solved prior to system implementation that, if feasible, 
guarantee the simultaneous satisfaction of all system requirements 
at runtime. 
Our contributions are:
\begin{itemize}
    \item A theoretical upper-bound for class $\K$ functions in an HOCBF, such that any class $\K$ function satisfying the bound validates that the HOCBF satisfies its definition (Theorem~\ref{thrm:altHOCBFdef}).
    \item A sequence of SOSPs that (i) generates the class $\K$ functions of a finite number of HOCBFs and (ii) guarantees the continued feasibility of the 
    sequence of online optimization problems for a system with CLFs and input constraints (Theorem~\ref{thrm:singleSynth}, Algorithm~\ref{alg:multiSynth}).
    \item Numerical simulations that demonstrate the practical implementation of the proposed framework for a system that has seven HOCBFs of relative degree two with a total of 14 unknown class $\K$ functions (Section~\ref{sec:Sims}).
\end{itemize}

 Existing work has established approaches for safe system verification by leveraging the connection between SOS programming and nonnegativity certificates
\cite{Clark_Verification,wang2018permissive,pond2023fast,Isaly,de2023synthesis,jiang2026verification,dai2024verification,schneeberger2023sos}.
In \cite{wang2018permissive,Isaly,zeng2021safety,pond2023fast,schneeberger2023sos,dai2024verification}, methods for system verification with CBFs are proposed.
However the techniques used for CBFs cannot be generalized in a straight-forward fashion to HOCBFs.
In \cite{Clark_Verification}, SOS procedures are proposed for systems with multiple HOCBFs. 
However, system actuation limits and stability requirements are not accounted for. 
In \cite{de2023synthesis}, 
SOS techniques are used to construct class~$\K$ functions for a single HOCBF. 
However, that work does not account for multiple HOCBFs, nor the inclusion of a CLF.

In \cite{jiang2026verification}, a complementary approach to that proposed in this paper is presented, where linear class $\K$ functions are fixed a priori in order to search for both HOCBF(s) and a CLF that belong to a safe region of the state space while obeying input constraints. 
In our contribution, we suppose that we are given
(i) a collection of functions that encode the system's safe region and (ii) a function that encodes the system's stability goals, and this problem formulation is in line with that of~\cite{xiao2021HOCBFs,mestres2024distributed,de2023synthesis}. 
Our method returns a collection of class $\K$ functions so that the functions in point (i) are valid HOCBFs, the function from point (ii) is a valid control Lyapunov-like function, and a continuous input exists that simultaneously satisfies these constraints.

The remainder of this paper is organized as follows. 
Section~\ref{sec:Prelims} gives background and a problem statement.
Section~\ref{sec:Bounds} formulates bounds on the class $\K$ functions of an HOCBF.
Section~\ref{sec:Synthesis} develops a sequence of SOSPs for safe system verification.
Section~\ref{sec:Sims} presents numerical simulations, 
and 
Section~\ref{sec:Conclusion} concludes.

\textbf{Notation} Let~$\mathbb{N}$ denote the positive integers.
For~$N \in \mathbb{N}$ we define $[N]:=\{1,\dots,N\}$.
The Lie derivative of a function $f$ with respect to a function $g$ is expressed as $L_gf(x) :=\frac{\partial f(x)}{\partial x}g(x)$.
Let $\R$ denote the reals and let $\R_+$ denote the nonnegative reals.
We use $\R[x]$ to denote the set of all real-valued, scalar polynomial functions and $\R^{m\times n}[x]$ to denote the set of real, matrix-valued polynomial functions. 
If $p\in\R^{m\times n}[x]$, then the entry $p_{ij}\in\R[x]$ for all $i\in[m]$ and $j\in[n]$.
We use $p\in\R^m[x]$ as shorthand for $p\in\R^{m\times 1}[x]$.
The set of all real-valued, scalar SOS polynomials is denoted $\Sigma[x]$, where
$s\in\Sigma[x]$ implies that $s(x)=\sum_ip_i^2(x)$ for some $p_i\in\R[x]$.
We use $\Sigma^{m\times n}[x]$ to represent the set of all real, matrix-valued polynomial functions, where $s\in\Sigma^{m\times n}[x]$ implies that $s_{ij}\in\Sigma[x]$ for all $i\in[m]$ and $j\in[n]$.
For shorthand, we use $\Sigma^m[x]$ in place of $\Sigma^{m\times 1}[x]$.

\section{Preliminaries and Problem Statements}
\label{sec:Prelims}

Consider the control-affine system
\begin{equation} \label{eq:dynamics}
    \dot{x} = f(x) + g(x) u,
\end{equation}
where the state is $x\in\X \subset \R^n$ and the input is $u\in\U \subset \R^m$. The set
$\X=\{x\in\R^n\mid h(x)\geq 0\}$, where $h\in\R^p[x]$ imposes $p\in\mathbb{N}$ constraints on the state.
The set $\U=\{u\in\R^m\mid c(u)\geq 0\}$, where $c\in\R^q[x]$  imposes $q\in\mathbb{N}$ constraints on the input.
The functions $f:\R^n\rightarrow\R^n$ and $g:\R^n\rightarrow\R^{n\times m}$ are locally Lipschitz everywhere. 
Then from any~$x_0 \in \X$, for some~$\tau_{max} > 0$ there is a unique solution~$x(t)$
to~\eqref{eq:dynamics} for $t \in I(x_0) := [0,\tau_{\max})$ \cite{ames2019control}.

\begin{assumption}\label{ass:fgpoly}
    The functions~$f$ and~$g$ in \eqref{eq:dynamics}
    are polynomials 
    and the sets $\X$ and $\U$ are compact.
\end{assumption}

We will enforce safety by making a given safe set $C$ forward-invariant. 

\begin{definition}[\cite{blanchini2008set} pg. 121]\label{def:forwardInvariant}
    A set $C\subset \R^n$ is \emph{forward invariant} with respect to the system \eqref{eq:dynamics} if every solution~$x$ with $x(0) = x_0\in C$ satisfies $x(t)\in C$ for all $t\in I(x_0)$. \hfill $\triangle$
\end{definition}

We will enforce safety using HOCBFs,
which are CBFs with non-trivial relative degree. 

\begin{definition}
    For the dynamical system \eqref{eq:dynamics} and an $r$-times continuously differentiable function $\psi_0: \R^n \rightarrow\R$, the \emph{relative degree} $r\in\mathbb{N}$ is the number of times that $\psi_0$ must be differentiated along the dynamics until the input $u\in\R^m$ uniformly appears. \hfill $\triangle$
\end{definition}

For a function $\psi_0$, 
we define a
sequence of high-degree functions 
 for all $i\in[r]$
as $ \psi_i(x) =\dot{\psi}_{i-1}(x)+\alpha_i(\psi_{i-1}(x))$, 
where $\alpha_i:[0,a)\rightarrow\R_+$ for $i\in[r]$ are user-specified class $\K$ functions \cite[Definition 4.1]{khalil2015nonlinear}.
For~$\psi_r$, we abuse notation by writing $\psi_r(x,u)$ instead of $\psi_r(x)$, which we do to emphasize that~$u$ explicitly appears in the expression for~$\psi_r$. 
The function $\psi_r$ can be expanded as $\psi_r(x,u) = L_f^r\psi_0(x) + L_gL_f^{r-1}\psi_0(x)u
+\sum_{i=1}^{r-1}L_f^i(\alpha_{r-i}(\psi_{r-i-1}(x)))+\alpha_r(\psi_{r-1}(x))$,
where $\psi_{r-1}$ is a function of every $\alpha_i$ with $i\leq r-1$ and $\psi_{r-i-1}$ is a function of every $\alpha_i$ with $i\leq r-i-1$.

\begin{definition}[\cite{xiao2021HOCBFs} Definition 8]\label{def:HOCBF}
    An $r$-times continuously differentiable function $\psi_0:\R^n\rightarrow \R$ is a relative degree $r$ HOCBF for the dynamical system \eqref{eq:dynamics} if there exist $r$ locally Lipschitz continuous class $\K$ functions $\alpha_i:[0,a)\rightarrow \R_+$ for $i\in[r]$ such that $\sup\limits_{u\in\U}[\psi_r(x,u) ]\geq 0$ for all $x\in \C_r$. \hfill $\triangle$
\end{definition}

We define a sequence of $r$ companion sets 
associated with~$\psi_0$ 
as $\C_i = \{x\in \R^n\mid \psi_{i-1}(x)\geq 0\}$ for all $i\in[r]$.
Note that $\C_r\subseteq\C_{r-1}\subseteq\dots\subseteq \C_1$
by construction.
To illustrate this nesting property, if $x\in\C_r$, then $\psi_{r-1}(x) \geq0$, where $\psi_{r-1}(x) = \dot{\psi}_{r-2}(x) + \alpha_{r-1}(\psi_{r-2}(x))$. 
Since $\alpha_{r-1}$ is only defined on $[0,a)$, 
we must have $\psi_{r-2}(x) \geq 0$.
For HOCBFs, the safe set is defined as $C=\C_1\cap\dots\cap\C_r$,
and~$C=\C_r$ 
due to the nesting property of the companion sets. 

\begin{assumption}\label{assum:companions}
    For all~$i \in [r]$, 
    the companion sets $\C_i$ 
     are nonempty and compact.
\end{assumption}




We define the admissible input set for the system in~\eqref{eq:dynamics} 
as~$\Psi_r(x) = \{u\in\U\mid \psi_r(x,u) \geq 0\}$.
The following lemma connects an HOCBF to the forward invariance of $C$.

\begin{lemma}[\cite{xiao2021HOCBFs} Theorem 4]\label{lem:hocbf}
    Consider the dynamical system \eqref{eq:dynamics} and an $r$-times continuously differentiable function $\psi_0:\R^n\rightarrow \R$. If $\psi_0$ is a relative degree $r$ HOCBF satisfying Definition \ref{def:HOCBF}, then any Lipschitz continuous controller $u:[0,\infty) \rightarrow \U$ such that $u(t) \in\Psi_r(x(t))$ for all $t\geq 0$ renders the set $C=\C_r$ forward invariant for the system.
\end{lemma}

We consider systems with $J\in\mathbb{N}$ safety constraints,
where the~$j^{th}$ safe set is encoded by an $r_j$-times continuously differentiable function $\psi^j_0$.
For all $i\in[r_j]$ and $j\in [J]$, the companion sets are~$\C^j_i=\{x\in\R^n\mid \psi^j_{i-1}(x)\geq 0\}$.
The $j^{\text{th}}$ safe set is denoted $C^j=\C^j_{r_j}$.
Since each $\psi^j_0$ and the dynamics are known, the relative degrees $r_j$ can be straightforwardly found by differentiating $\psi^j_0$ along the dynamics.
For complicated examples, this can be performed by a software program, such as MATLAB's Symbolic Math Toolbox.

For real-time implementation, HOCBFs are commonly incorporated as constraints 
in a sequential quadratic program (SQP), where time is discretized into the set $T=\{0,\Delta t,2\Delta t,\dots\}$ and a new QP is solved at each $t\in T$ to 
compute the input that is 
held constant over the interval $[t,t+\Delta t)$. 
We consider SQPs of the form 
\begin{subequations}\label{eq:SQP}
    \begin{align}
        &u^\star(t) = \argmin_{u(t),\rho(t)} ||u(t) - \hat{u}(t)||^2+\rho(t)^2 \label{eq:SQPobj}
         \\ &\textnormal{ subject to }\; \psi^j_{r_j}(x(t),u(t)) \geq 0 \quad \text{for all }j\in[J] \label{eq:SQPpsi1}
        \\ & L_fV(x(t)) + L_gV(x(t))u(t) + \gamma (V(x(t)))\leq \rho(t) \label{eq:SQPV}
        \\ &\qquad\qquad\quad c(u(t))\geq 0. \label{eq:SQPu}
    \end{align}
\end{subequations}
The objective in~\eqref{eq:SQPobj} depends on a nominal input $\hat{u}(t)\in\R^m$, which 
can be user-specified or found by computing an optimal input without
safety constraints.
The $J$ constraints in \eqref{eq:SQPpsi1} are the implementation of Lemma \ref{lem:hocbf}, and each one individually enforces 
the forward invariance of $C^j$ for a $j\in[J]$.
The constraint \eqref{eq:SQPu} enforces $u(t)\in\U$ for all $t\in T$.
In the SQP, it is common to relax a CLF constraint \eqref{eq:SQPV}
with a variable $\rho(t)\geq 0$ to prioritize safety over stability~\cite{ames2019control,xiao2021HOCBFs}. 
Ideally, $\rho^\star(t)=0$ for all $t\in T$. 
This condition is referred to as ``safe stabilization'' \cite{wang2018permissive,ames2019control}.
The choice of sampling frequency is a system-specific implementation decision.

In this work, we propose a technique to determine \textit{a priori} if the real-time implementation of safe stabilization is possible for a system.
To do so, we use control-Lyapunov like functions (CLlFs), which use
the level surfaces of a CLF to characterize a system's stability behavior.

\begin{definition}[\cite{blanchini2008set}]\label{def:CLlF}
    A locally Lipschitz function $V:\R^n\rightarrow \R$ is a \emph{control Lyapunov-like function} (CLlF) for the system in~\eqref{eq:dynamics} if it satisfies~$\min\limits_{u\in\U}[L_fV(x) +L_gV(x)u]\leq \rho,$ for all $x\in\X$
    with $\rho=0$.
    \hfill $\triangle$
\end{definition}


In many of the optimization problems we present, we minimize $\rho \geq 0$ subject to 
the constraint $L_fV(x)+L_gV(x)u\leq \rho$.
If $\rho^\star=0$ is returned, then it is known that $V$ is a CLlF for the system satisfying Definition \ref{def:CLlF}.
If $\rho^\star>0$ is returned, then it is not known whether $V$ is  a CLlF for the system.

Let $S$ be a finite collection
$S=\{v_1,\dots,v_k;w_1,\dots,w_s\}$, with $v_i,w_j\in\R[x]$ for all $i\in[k]$ and $j\in[s]$, the semialgebraic set generated by $S$ is $K_S=\{x\in\R^n\mid v_1(x)\geq0,\dots,v_k(x)\geq0, w_1(x) =0,\dots, w_s(x) =0\}$.

\begin{definition}[\cite{powers2021certificates}]\label{def:quadraticModule}
The quadratic module generated by $S\subseteq\R[x]$ is defined as
\begin{multline}\label{eq:quadraticModule}
   Q_S = \Big\{q\in\R[x] :  q(x)  = s_0(x) \\+ \sum_{i=1}^k s_i(x)v_i(x) + \sum_{j=1}^ s p_j(x)w_j(x)\Big\},
\end{multline}
where $s_i\in\Sigma[x]$ for all $i\in\{0\}\cup [k]$ and $p_j\in\R[x]$ for all $j\in[s]$. \hfill $\triangle$
\end{definition}

The following lemma is a fundamental nonnegativity result in real
algebraic geometry. 

\begin{lemma}[\cite{powers2021certificates}]\label{lem:nonneg}
    Suppose $K_S\not=\emptyset$. If there exists an explicit representation of a polynomial $g\in Q_S$ in the form of \eqref{eq:quadraticModule}, then $g\geq 0$ on the semialgebraic set $K_S$.
\end{lemma}

Recall that the HOCBF definition in Definition \ref{def:HOCBF} 
requires the nonnegativity of a polynomial $\psi_r$ on a semialgebraic set $\C_r$.
The following assumption is required in order to implement Lemma \ref{lem:nonneg} with SOS programming.

\begin{assumption}\label{ass:psipoly}
    The high-degree function $\psi^j_i\in\R[x]$ for all $i\in[r_j]$ and $j\in[J]$ and the CLlF $V\in\R[x]$ are scalar-valued polynomial functions.
\end{assumption}

\subsection{Problem Statement}

The goal of this work is to provide an \textit{a priori} certificate that ensures the SQP 
in~\eqref{eq:SQP} is feasible for all $t\in T$.
The safe operating region of a system with~$J$ HOCBFs is defined as $C_\sys=C^1\cap \dots \cap C^J$.


\begin{definition}\label{def:guarantee}
A dynamical system (1) has a \emph{guarantee of continued feasibility} for the real-time implementation of the SQP in \eqref{eq:SQP} if there exists a Lipschitz continuous input $u(x)$ for all $x\in C_\sys\not=\emptyset$ that simultaneously satisfies: 
    \begin{enumerate}
        \item The HOCBF definition 
        in Definition~\ref{def:HOCBF}
        and the forward invariance lemma in 
        Lemma~\ref{lem:hocbf} 
        for all $j\in[J]$,
        \item The CLlF definition in Definition~\ref{def:CLlF}
        for some $\rho\geq0$,
        \item The actuation constraints $u(x)\in\U$. \hfill $\triangle$
    \end{enumerate}
\end{definition}

If a guarantee of continued feasibility is established, 
then the SQP in \eqref{eq:SQP} is feasible and produces a continuous solution.
Provided that the sampling interval $\Delta t$ is sufficiently small, we solve 
\eqref{eq:SQP} via a sample-and-hold controller as in \cite{ames2016control,xiao2019control,galloway2015torque}.
We refer readers to \cite{gurriet2019realizable} for information regarding the relationship of continuous-time controllers and their discrete-time implementations with regard to CBFs.
We note that a guarantee of continued feasibility does not hold in general~\cite{xiao2021adaptive,garg2024advances,zeng2021safety}. 
Hence we seek to provide such guarantees by solving 
a sequence of SOSPs one time, prior to the real-time implementation with the SQP in~\eqref{eq:SQP}.

\begin{problem}\label{prob:synthesis}
    Suppose there are given (i)
    $J$ safe sets $C^j$ for $j \in [J]$, each defined by an $r_j$-times  continuously differentiable function $\psi^j_0:\R^n\rightarrow\R$, 
    and (ii) a CLlF $V:\R^n\rightarrow \R$.
    Then, for a dynamical system \eqref{eq:dynamics} with an input constraint set $\U$, determine a sequence of SOSPs that returns all class $\K$ functions $\{\alpha^j_i\}_{i\in[r_j]}$ for $j\in[J]$ such that the system has a guarantee of continued feasibility for the SQP \eqref{eq:SQP} in the sense of Definition \ref{def:guarantee}.
\end{problem}




\section{A Theoretical Bound on Class $\K$ Functions}
\label{sec:Bounds}


In practice class $\K$ functions are often chosen and assumed to be adequate, 
without explicit validation of Definition~\ref{def:HOCBF}.
However, if that definition is not satisfied for all $x\in \C_r$, then there may be a point in time at which 
the SQP \eqref{eq:SQP} is infeasible.
A fundamental challenge in verifying Definition~\ref{def:HOCBF} 
is that there are multiple class $\K$ functions to find for 
each HOCBF, and they are recursively dependent 
in the sense 
that $\alpha_k$ depends on~$\alpha_i$ for all $i < k$.

\begin{figure*}[htbp]
    \centering
    \includegraphics[width=.7\linewidth]{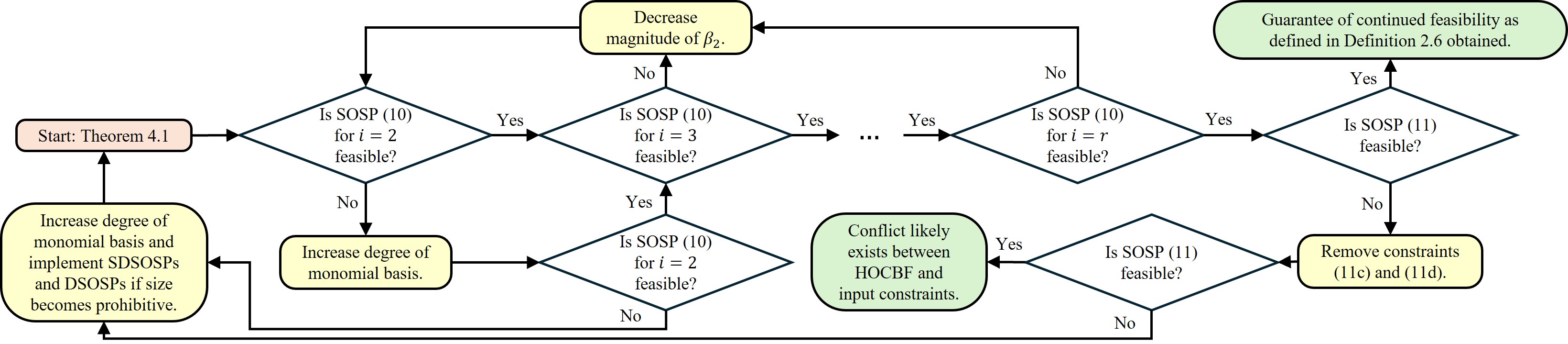}
    \caption{Recommended heuristic techniques for Theorem \ref{thrm:singleSynth}.}
    \label{fig:flowchart}
\end{figure*}

Now we present an alternate version of Definition \ref{def:HOCBF} and Lemma \ref{lem:hocbf} that do not explicitly depend on $\{\alpha_i\}_{i\in[r]}$.

\begin{theorem}\label{thrm:altHOCBFdef}
    Consider the system \eqref{eq:dynamics} and a safe set $C$ defined by an $r-$times continuously differentiable function $\psi_0:\R^n\rightarrow \R$.
    Let Assumptions \ref{ass:fgpoly}, \ref{assum:companions}, and \ref{ass:psipoly} hold.
    \begin{enumerate}[label=(\alph*)]
        \item Suppose that there exist polynomial functions $\beta_i\in\R[x]$ with $\beta_i(0)=0$ for all $i\in\{2,\dots,r\}$ such that
        \begin{align}
            \ddot{\psi}_{i-2}(x) \geq \beta'_{i-1}(\psi_{i-2}(x)) \; \text{for all } &x\in \C_{i-1} \label{eq:thrmpsi0}
            \\ \beta_{i-1}'(\psi_{i-2}(x)) >0 \; \text{for all } &x\in \text{int}(\C_{i-1}) \label{eq:thrmbeta1int}
            \\ \beta_{i-1}'(\psi_{i-2}(x))\geq 0 \; \text{for all }&x\in \partial \C_{i-1}\label{eq:thrmbeta1par}.
        \end{align}
        Then there exist class $\K$ functions $\alpha_i\in\R[x]$  
        bounded via
            $\alpha_{i-1}(\psi_{i-2}(x))\leq \sqrt{2\beta_{i-1}(\psi_{i-2}(x))}$
        for all $i \in [r] \backslash \{1\}$
        such that the safe set $C=\C_r$ is forward invariant for \eqref{eq:dynamics}.
        \item Suppose that there exists a Lipschitz continuous 
        input $u:[0,\infty)\rightarrow \U$ and a function $\alpha_r\in\R[x]$ such that
        \begin{align}
             \dot{\psi}_{r-1}(x,u)\geq -\alpha_r(\psi_{r-1}(x))  \; \text{for all } &x\in \C_r \label{eq:thrmpsir-1}
             \\ \alpha_r'(\psi_{r-1}(x)) >0 \; \text{for all }&x\in \text{int}(\C_r) \label{eq:thrmbetar-1}
             \\ \alpha_r(\psi_{r-1}(x)) = 0 \; \text{for all }&x\in \partial \C_r\label{eq:thrmbetar-1par}. 
        \end{align}
        Then $\alpha_r$ is a class $\K$ function such that~$u(t)\in\Psi_r(x(t))$
        for all $t\geq0$.
    \end{enumerate}
    
\end{theorem}

\emph{Proof: } See Appendix \ref{app:altHOCBFdef}. \hfill $\blacksquare$
SOSPs are solved via semidefinite programs (SDPs).
This class of programs requires
the decision variables to appear linearly.
The only high-degree function that depends on just one class $\K$ function is $\psi_1(x) = \dot{\psi}_0(x) + \alpha_1(\psi_0(x))$ because $\psi_0$ itself is not a function of any class $\K$ functions.
This observation motivates the ``bottom-up'' structure of Theorem \ref{thrm:altHOCBFdef}, in which we first generate $\alpha_1$ and use it to find $\alpha_2$ and so forth until we find $\alpha_r$.
This is in contrast to the ``top-down'' reasoning in Definition \ref{def:HOCBF} and Lemma \ref{lem:hocbf}, 
which does not readily admit an SOSP implementation. 
Indeed, the recursive dependence of the class~$\K$ functions 
prevents a direct extension of previous SOS-based CBF verification methods, since the 
linear appearance of decision variables implies that 
class $\K$ functions and the input cannot be solved for simultaneously.

Theorem \ref{thrm:altHOCBFdef} holds for any $\beta_i$ that satisfies conditions \eqref{eq:thrmbeta1int} and \eqref{eq:thrmbeta1par}.
It is sufficient for our purposes to express
$\beta_i\in\R[x]$ 
as~$\beta_i(\zeta) = b_{i,1}\zeta + \frac{1}{2} b_{i,2}\zeta^2 + \dots + \frac{1}{2m} b_{i,2m}\zeta^{2m}$ for $i\in[r]\backslash\{1\}$,
where $b_{i,j}\in\R_+$ are unknown coefficients and $m$ is a user-selected parameter. 
With this formulation, the derivative of $\beta_i$ is $\beta_i'(\zeta) = b_{i,1} + b_{i,2}\zeta + \dots + b_{i,2m}\zeta^{2m-1}$ for $i \in [r] \backslash \{1\}$. 
We use this formulation because $\beta_i(0)=0$, $\beta_i'(0)\geq 0$, and $\beta_i'(\zeta)>0$ for $\zeta\in[0,\bar{\zeta}_{i}]$, where $\bar{\zeta}_i=\max_{x\in\C_i}\psi_{i-1}(x)$, are enforced automatically, and 
therefore these conditions do not 
need to be imposed as additional constraints in the forthcoming SOSPs.
Similarly, for $\alpha_r\in\R[x]$, we use $\alpha_r(\zeta) = a_{r,1}\zeta + \frac{1}{2}a_{r,2}\zeta^2+\dots + \frac{1}{2m}a_{r,2m}\zeta^{2m},$
with $a_{r,j}\in\R_+$,
because $\alpha_r(0)=0$ and $\alpha'_r(\zeta)>0$ for all $\zeta\in(0,\bar{\zeta}_r)$ are enforced by definition. 

\section{Safe System Verification via SOS Programming}
\label{sec:Synthesis}

We develop a sequence of SOSPs that generates a collection of class~$\K$
functions so that $\psi_{r}$ satisfies the HOCBF definition in Definition \ref{def:HOCBF}.
Our method relies on SOS programming and is therefore only able to search for feasibility guarantees for polynomial controllers.
However, this polynomial limitation is counterbalanced by the result that
any continuous function can be uniformly approximated arbitrarily closely with a polynomial \cite{stone1948generalized}.
For this reason, we expect it to be unusual for our verification methods 
to fail due to a dynamic system \textit{only} admitting a non-polynomial safe controller.

\begin{theorem}\label{thrm:singleSynth}
    Consider the system \eqref{eq:dynamics} with actuation constraint set $\U$,
    a safe set $C$ defined by an $r$-times continuously differentiable function $\psi_0:\R^n\rightarrow\R$, and a CLlF $V:\R^n\rightarrow \R$. 
    Let Assumptions \ref{ass:fgpoly}, \ref{assum:companions}, and \ref{ass:psipoly} hold.
    The system has a guarantee of continued feasibility in the sense
    of Definition~\ref{def:guarantee} for the SQP in \eqref{eq:SQP} if there exist (i) a polynomial $u\in\R^{m}[x]$, (ii) functions $\beta_i\in\R[x]$ for $i\in\{2,\dots,r\}$, and (iii) $\alpha_r\in\R[x]$ that solves the following sequence of SOSPs:
    \begin{enumerate}[label=(\alph*)]
        \item For $i\in[r]\backslash\{1\}$, beginning with $i=2$, solve the following feasibility program to determine $\beta_{i-1}^\star\in\R[x]$: 
        \begin{subequations}\label{eq:SSSDP1}
            \begin{align}
                &\underset{\substack{\beta_{i-1}\in\R[x]\\s^{i-1}_0, s^{i-1}_1\in\Sigma[x], s^{i-1}_2\in\Sigma^p[x]}}{\text{minimize}} \; 0 \label{eq:SSSobjfun}
            \\ &\text{subject to } \ddot{\psi}_{i-2}(x) = \beta_{i-1}'(\psi_{i-2}(x)) + s^{i-1}_0(x)
            \\ &\quad\quad+ s^{i-1}_1(x) \psi_{i-2}(x) +(s^{i-1}_2(x))^T h(x). \label{eq:SSi}
            \end{align}
        \end{subequations}
        \item Solve the following optimization problem to determine $\alpha_r^\star\in\R[x]$ and the stability measure $\rho^\star\in\R_+$:
        \begin{subequations}\label{eq:SSSDP2}
            \begin{align}
                &\underset{\substack{\rho\in\R_+, u\in\R^m[x], \alpha_r\in\R[x] \\
                s^r_0,s^r_1,\hat{s}_0,\hat{s}_1\in\Sigma[x] \\ 
                s^r_2,\hat{s}_2\in\Sigma^p[x],  
                \tilde{s}_0,\tilde{s}_1\in\Sigma^q[x], \tilde{s}_2\in\Sigma^{p\times q}[x]}}{\text{minimize}} \; \rho \label{eq:SSSobjfun2}
            \\ &\text{subject to } \dot{\psi}_{r-1}(x,u(x)) = -\alpha_r(\psi_{r-1}(x)) +s^r_0(x)
            \\  &\quad\quad\quad\quad\quad + s^r_1(x) \psi_{r-1}(x) +(s^r_2(x))^Th(x)\label{eq:SSr}
            \\ -&L_fV(x) - L_gV(x)u(x) = -\rho + \hat{s}_0(x) 
            \\ &\quad\quad\quad\quad\quad+ \hat{s}_1(x) \psi_{r-1}(x) + (\hat{s}_2(x))^Th(x) \label{eq:SSV}
            \\ &c(u(x)) = \tilde{s}_0(x) + \tilde{s}_1(x) \psi_{r-1}(x)
            \\ &\quad\quad\quad\quad\quad+(\tilde{s}_2(x))^Th(x). \label{eq:SSU}
            \end{align}
        \end{subequations}
    \end{enumerate}
\end{theorem}
\noindent \emph{Proof: } See Appendix \ref{app:singleSynth}. \hfill $\blacksquare$ 
The SOSPs in Theorem \ref{thrm:singleSynth}
only need to be feasible, in the sense that a solution
to them must exist. 
For this reason, the objective function can change across implementations and system constraints can be added to and removed  
based on the system requirements.



It is possible for the sequence of SOSPs in Theorem \ref{thrm:singleSynth} to be infeasible.
Possible causes include
incompatibility between the input constraints and the barrier function,
incompatible functions~$\beta_i$ for $i\in[r]$,
or an inadequate degree of the monomial basis chosen for the SOSPs. 

If infeasibility is encountered in one of the (a) SOSPs, then we recommend re-running the sequence, but dividing the first returned $\beta_i$ function by a constant to decrease its magnitude. Typically we divide by~$10$ when this occurs.
We have observed that if $\beta_i$ takes large values, especially early in the sequence with lower values of the index $i$, 
then that can negatively impact the feasibility of the SOSPs that follow.
If infeasibility is encountered in the final (b) SOSP, we recommend re-running the SOSP with constraints \eqref{eq:SSV} and \eqref{eq:SSU} removed. 
If a solution is then reached, it is likely that there is a conflict between the CBF and the input
constraints, which can be remedied by changing 
one or both of these parts of the problem formulation to produce a feasible problem.



Another aspect that can impact feasibility is an inadequate degree of the monomial basis for the SOS programs. 
As the degree of the monomial basis grows in an SOSP, its feasible region grows, 
but problems become more computationally complex because 
the complexity of an SOSP grows exponentially quickly in the degree of the basis.

Computationally simpler alternatives to SOSPs include diagonally-dominant sum-of-squares
(DSOS) 
and scaled diagonally-dominant sum-of-squares
(SDSOS) programs, which are solved via linear programming and second-order cone programming, respectively \cite{ahmadi2019dsos,megretski}. 
However, a tradeoff also exists when using these alternative programs, since the feasible region of a DSOS program 
inner-approximates
the feasible region of an SDSOS program, which is then an inner-approximation of the feasible region of an SOS program \cite{majumdar2014control,ahmadi2015sum}.    
Therefore, these approaches may fail to solve verification problems that would be 
solved by an SOS approach.
    
Nonetheless, 
when increasing the degree of the monomial basis in practice, it can be fruitful to use DSOS or SDSOS programs when the computational burden of the SOSPs becomes prohibitive~\cite{pond2023fast}.
Our results do not require a commitment to a specific implementation choice,
    and we formulate problems to allow any approach to be used in practice.
Figure \ref{fig:flowchart} provides a flowchart of the heuristic techniques we recommend using if infeasibility occurs in the application of Theorem \ref{thrm:singleSynth}.



We now consider systems with multiple HOCBFs. 
Example uses include multi-agent systems 
in which each robot has a barrier function for each other robot 
in the system \cite{mestres2024distributed,tan2021distributed}.

These systems must satisfy $x(t)\in C_\sys=C^1\cap\dots\cap C^J$ for all $t \geq 0$, but
we do not require each~$C^j$ to be forward invariant. 
We next present variations of Definition \ref{def:HOCBF} and Lemma \ref{lem:hocbf} 
for the multi-HOCBF case. 

\begin{definition}\label{def:multiHOCBF}
    An $r_j$-times continuously differentiable function $\psi^j_0:\R^n\rightarrow \R$ is a relative degree $r_j$ \emph{HOCBF for a multi-HOCBF system} \eqref{eq:dynamics} with $J$ total HOCBFs if there exist $r_j$ class $\K$ functions $\alpha^j_i:[0,a)\rightarrow \R_+$ 
    with~$i \in [r_j]$ 
    such that $\sup_{u\in\U} \psi^j_{r_j}(x,u) \geq 0 \quad \text{for all }x\in C_\sys.$ \hfill $\triangle$
\end{definition}

Let $\bar{r}=\max_{j\in[J]}r_j$. The modified admissible set is defined as~$\Psi_{\bar{r}}(x) = \{u\in\U\mid \psi^1_{r_1}(x,u) \geq 0,\dots, \psi^J_{r_J}(x,u) \geq 0\}.$
\begin{lemma}\label{lem:multiHOCBF}
    Consider the dynamical system \eqref{eq:dynamics} and $J$ HOCBFs of the form $\psi^j_0:\R^n\rightarrow \R$ for $j\in[J]$.
    If $\psi^j_0$ is a relative degree $r_j$ HOCBF satisfying Definition \ref{def:multiHOCBF} for all $j\in[J]$, then any Lipschitz continuous controller $u:[0,\infty)\rightarrow \U$ such that $u(t)\in\Psi_{\bar{r}}(x(t))$ for all $t\geq 0$ renders the system safe set $C_\sys$ forward invariant.
\end{lemma}

\noindent \emph{Proof: } See Appendix \ref{app:multiHOCBF}. \hfill $\blacksquare$

To verify system safety, we iteratively implement the sequence of SOSPs in Theorem \ref{thrm:singleSynth} for each of the safety constraints and incorporate
the previously verified $j^{th}$ HOCBF into the verification procedure for 
the $(j+1)^{th}$ candidate. 
Algorithm~\ref{alg:multiSynth} formalizes this procedure. 

As in the proof of Theorem \ref{thrm:singleSynth}, for the SOSP \eqref{eq:SSSDP1}, the quadratic module used is $Q_{S_{i-1}}=\{q\in\R[x]\mid q(x) = s^{i-1}_0(x) + s^{i-1}_1(x)\psi_{i-2}(x) + (s^{i-1}_2(x))^Th(x)\}$, where $s^{i-1}_0,s^{i-1}_1\in\Sigma[x]$ and $s^{i-1}_2\in\Sigma^p[x]$ are arbitrary.
For the SOSP in \eqref{eq:SSSDP2}, the quadratic module used is $Q_{S_{r}}=\{q\in\R[x]\mid q(x) = s^{r}_0(x) + s^{r}_1(x)\psi_{i-1}(x) + (s^{r}_2(x))^Th(x)\}$, where $s^{r}_0,s^{r}_1\in\Sigma[x]$ and $s^{r}_2\in\Sigma^p[x]$ are arbitrary, and the $q-$dimensional quadratic module $Q_{u} = \{q_u\in\R^q[x]\mid q_u(x) = \tilde{s}_0(x) + \tilde{s}_1(x)\psi_{r-1}(x) + (\tilde{s}_2(x))^Th(x)\}$, with $\tilde{s}_0,\tilde{s}_1\in\Sigma^q[x]$ and $\tilde{s}_2\in\Sigma^{p\times q}[x]$ arbitrary.

\begin{algorithm}
\caption{System Verification through Class $\K$ Generation }\label{alg:multiSynth}
    \If{$J>1$}{
    \For{$j\in[J]$}{
    \For{$i\in\{2,\dots,r_j\}$}{
    Set $S'_{i-1}=\{\psi^1_{i-2},\dots,\psi^j_{i-2},h_1,\dots,h_p\}$\\
    Generate the one-dimensional quadratic module $Q_{S_{i-1}'}$ defined by $S_{i-1}'$ with \eqref{eq:quadraticModule} \\
    Solve \eqref{eq:SSSDP1} for $\alpha_{i-1}^{j,\star}\in\R[x]$ using $Q_{S_{i-1}'}$ instead of $Q_{S_{i-1}}$
    }
    Set $S'_{r_j} = \{\psi^1_{r_1-1},\dots,\psi^j_{r_j-1},h_1,\dots,h_p\}$\\
    Generate the one-dimensional quadratic module $Q_{S'_{r_j}}$ defined by $S_{r_j}'$ with \eqref{eq:quadraticModule} \\
    Generate the $q-$dimensional quadratic module $Q_{u'}$ defined by $S_{r_j}'$ with \eqref{eq:quadraticModule}\\
    Solve \eqref{eq:SSSDP2} for $\alpha^{j,\star}_{r_j}\in\R[x]$ using $Q_{S'_{r_j}}$ instead of $Q_{S_{r_j}}$ and $Q_{u'}$ instead of $Q_{u}$}
    }
\end{algorithm}

The iterative nature of Algorithm~\ref{alg:multiSynth}
ensures that each new HOCBF preserves the validity of previously
verified HOCBFs. 
Algorithm~\ref{alg:multiSynth} requires solving $J\sum_{j=1}^Jr_j$ SOSPs.
The final SOSP in the process, corresponding to candidate $j=J$ and SOSP $i=r_J$, guarantees the continued feasibility of the system \eqref{eq:dynamics}
in the sense of Definition \ref{def:guarantee}.

\section{Simulations}
\label{sec:Sims}
\begin{figure*}
\centering
    \includegraphics[width=.85\textwidth]{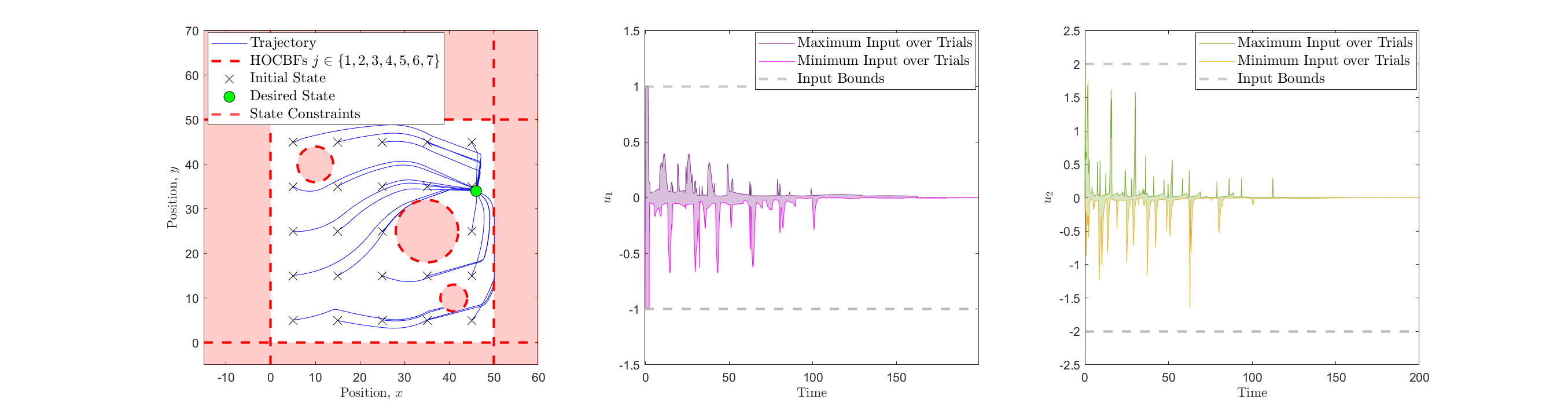}
    \caption{
    SQP results for the seven verified HOCBFs. All trajectories stay within the system's safe set $C_\sys = C^1\cap\dots\cap C^7$, visually outlined by the dashed red lines. 
    The left-most plot shows the trajectories from several initial conditions. 
    The center and right plots display envelopes that contain the inputs $u_1(t)$ and $u_2(t)$,
    respectively, for each of the trials.}
    \label{fig:synthSims}
\end{figure*}

In this section, we use Algorithm \ref{alg:multiSynth} to generate the class $\K$ 
functions needed to implement seven safety constraints and verify the continued feasibility of a unicycle system.
The dynamics are $\dot z= \begin{bmatrix} v \cos\theta & v \sin\theta & 0 & 0\end{bmatrix}^T+\begin{bmatrix} 0 & 0 & u_1 & u_2\end{bmatrix}^T$,
with $z=\begin{bmatrix} x & y & v & \theta \end{bmatrix}^T\in\X$ and $u= \begin{bmatrix} u_1 & u_2 \end{bmatrix}^T\in \U$, where $\X = [0,50] \times [0, 50] \times [0,2] \times [-\pi/2, \pi/2]$ and $\U = [-1, 1]\times [-2, 2]$ \cite{xiao2021HOCBFs}.
We use the polynomial approximations 
$\cos\theta = 1- \frac{\theta^2}{2}$ and $\sin \theta = \theta - \frac{\theta^3}{6}$.

The system we consider has $J = 7$ safe regions encoded in terms of the state variables $x$ and $y$.
For each~$j \in [J]$, $\psi^j_0$ has relative degree two and 
has two associated
unknown class $\K$ functions $\alpha^j_1$ and~$\alpha^j_2$. 
We use three CLFs to direct the state to a desired $(x,y)$-position and velocity~$v$.

For $j\in[3]$, the safe sets are defined by $\psi^j_0(z) = (x-x_0)^2 + (y-y_0)^2 - R_j^2$, 
and they ensure that the system's 
state stays out of three circles 
in the plane.
The other four safety constraints with indices $j\in\{4,5,6,7\}$ 
enforce the state constraints that define the feasible region $\X$. 
Three CLlFs are used to guide the state to $[x_d,y_d,z_d] = [46, 34, 0.2]$, and
our verification procedure can accommodate more than one CLlF by incorporating three 
copies of constraint \eqref{eq:SSV} in each SOSP.

For each candidate $j\in[J]$, we use Theorem \ref{thrm:singleSynth} to find $\{\alpha^{j,\star}_i\}_{i=1,2}$. 
For the SOSP corresponding to $i=1$ given by \eqref{eq:SSSDP1}, we use $m^1_j=1/2$ for all $j\in[J]$ to define $\beta_1^{',j}\in\R[z]$.
Therefore, the function 
$\beta_1^{j,\star}\in\R[z]$
takes the form $\beta_1^{j,\star}(\psi^j_0(z)) = b_{1}^{j,\star}\psi^j_0(z).$
The class $\K$ functions must satisfy $ \alpha^{j,\star}_1(\psi_{0}^j(z))\leq \big(2\beta^{j,\star}_1(\psi^j_0(z))\big)^{1/2}
= a^{j,\star}_1 \big(\psi^j_0(z)\big)^{1/2}$
for all $z\in\C^j_1$,
where $a^{j,\star}_1 = \big(2b^{j,\star}_1\big)^{1/2}$.
However, in the following SOSP \eqref{eq:SSSDP2} that generates $\alpha^{j,\star}_2$, we must have a polynomial form of $\alpha^{j,\star}_1$, which we denote
by $\tilde{\alpha}^{j,\star}_1$.
For this we use $\tilde{\alpha}^{j,\star}_1(\psi^j_0(z)) = \eta^{j,\star}_1 \psi^j_0(z)$
for all $j\in[J]$ to satisfy the upper-bound $\tilde{\alpha}^{j,\star}_1(\psi^j_0(z))\leq 
\big(2\beta^{j,\star}_1(\psi^j_0(z))\big)^{1/2}$.
Here we have used $\eta^{j,\star}_1 = \sfrac{\big(\beta^{j,\star}_1(\bar{\zeta}^j_0)\big)^{1/2}}{\bar{\zeta}^j_0}$ and $\bar{\zeta}^j_0=\max_{z\in \C^j_1}\psi^{j}_0(z)$.
For the $i=2$ SOSP given by \eqref{eq:SSSDP2}, we select $m^2_j = 1/2$ for all $j\in[J]$ to define $\alpha_2^j\in\R[z]$.
Therefore, the class $\K$ function is 
$
    \alpha^{j,\star}_2(\psi^j_1(z)) = a^{j,\star}_2\psi^j_1(z). 
$
We also have three measures of safe stabilization $\rho_x,\rho_y,\rho_v\in\R_+$ for the CLlFs.

These results were generated in MATLAB using the SOSTOOLS toolbox\footnote{Code for these simulations is available online at: https://github.com/epond3/HOCBFVerificationSimulations.git} \cite{prajna2002introducing}.
Table \ref{tab:classK} lists the generated class $\K$ parameters.
Figure \ref{fig:synthSims} shows successful SQP implementation over multiple initial conditions, with trajectories remaining in $C_\text{sys}$.
The nonzero values of $\rho_x,\rho_y,\rho_v$ indicate that safe stabilization could not be certified, and therefore CLF slack variables were used.

\begin{table}[h!]
\centering
\caption{Class $\K$ function parameters $a^{j\star}_i$ used in the SQP.}
\label{tab:classK}
\begin{tabular}{|c | c | c | c | c | c | c | c|} 
 \hline
 \backslashbox{$i$}{$j$} & $1$ & $2$ & $3$ & $4$ & $5$ & $6$ & $7$ \\ [0.5ex] 
 \hline\hline
  \rule{0pt}{3ex}  1 & 0.63 & 8.3 & 0.17 & 1.7 & 1.7 &  0.10 & 0.13  \\  
  \hline
  \rule{0pt}{3ex} 2 & 1.1 & 12000 & 1.0 & 0.60 & 1.0 & 7.3 & 1.2  \\ [1ex] 
 \hline
\end{tabular}
\end{table}

\begin{table}[h!]
\centering
\caption{Computation times for Algorithm~\ref{alg:multiSynth}.}
\label{tab:compTime}
\begin{tabular}{|c||c|c|c|c|c|c|c|} 
 \hline
 \shortstack{Number of \\ HOCBFs} & 1 & 2 & 3 & 4 & 5 & 6 & 7 \\ 
 \hline
  \shortstack{Computation\\  Time (s)} & 127 & 247 & 397 & 561 & 720 & 889 & 1053 \\
 \hline
\end{tabular}
\end{table}


Table \ref{tab:compTime} reports cumulative computation time 
for each value of the counter~$j$
when running the 
outer \texttt{for}
loop in Algorithm~\ref{alg:multiSynth}.
These times are equal to 
the total elapsed time after each execution of line 10 in the algorithm block, and they
show a roughly linear relationship between the number of HOCBFs and computation time of Algorithm~\ref{alg:multiSynth}.

\section{Conclusion}
\label{sec:Conclusion}
This paper presented techniques for 
generating class~$\K$ functions that ensure
the continued feasibility of systems with multiple 
HOCBFs. 
In future work, these methods will be extended to procedures for distributed safe system verification.

\section*{References}
\bibliographystyle{ieeetr}
\bibliography{bib}

\appendix

\begin{lemma}[\cite{blanchini2008set}]\label{lem:Nagumos}
    Consider the system \eqref{eq:dynamics} and assume that for each $x_0\in\X$ the system admits a unique solution defined for all $t\in I(x_0)$. The set $C$ is forward invariant for \eqref{eq:dynamics} if and only if $\dot{x}\in T_C(x)$ for all $x\in \partial C$. \hfill $\blacksquare$
\end{lemma}


\begin{lemma}\label{lem:classKprime}
    Consider a class $\K$ function $\alpha:[0,a)\rightarrow \R_+$. Then
    $\alpha'(0)\alpha(0)=0$ and $\alpha'(\zeta)\alpha(\zeta)>0$ for all $\zeta\in(0,\infty)$.
\end{lemma}
\begin{proof}
    Since $\alpha(0)=0$ we have $\alpha'(0)\alpha(0)=0$.
    By definition we have~$\alpha'(\zeta)>0$ for all $\zeta\in(0,a)$. 
    Since $\alpha(\zeta)>0$ for all $\zeta\in(0,a)$, the result follows. 
\end{proof}

\subsection{Supporting Results}
The following corollary applies Nagumo's theorem in Lemma \ref{lem:Nagumos} to an intersection set $\cap_{i=1}^k\C_i$.

\begin{corollary}\label{corr:tangentCone}
    Consider an $r$-times continuously differentiable function
    $\psi_0:\R^n\rightarrow\R$ with companion sets $\C_i$ for $i\in[r]$. 
    The set $\cap_{i=1}^k\C_i$ is forward invariant for \eqref{eq:dynamics} with $k\leq r$
    if and only if $\dot{\psi}_{k-1}(x) \geq 0 \quad \text{for all }x\in \partial \C_k.$
\end{corollary}
\begin{proof}
Consider the tangent cone condition in Lemma \ref{lem:Nagumos}.
The set $\cap_{i=1}^k\C_i$ is forward invariant for the system if and only if $\dot{x}\in T_{\cap\C_i}(x)$ for all $x\in \partial (\C_k\cap\dots\cap \C_1)$, where
    $T_{\cap \C_i}(x) = \{z\in\R^n \mid \nabla \psi_{i-1}(x)^Tz\geq 0 \; \text{for all } i\in \text{Act}(x) \}$,
    where $\text{Act}(x)=\{i\leq k\mid \psi_{i-1}(x) = 0\}$ is the 
    set of indices of all active constraints at $x\in\X$.
    This condition is equivalently stated with the following $k$ constraints: $\dot{\psi}_{i-1}(x)\geq 0$ for all 
    $x\in\partial \C_i\cap\left( \cap_{j=1}^k\C_j\right)$ for every $i\in[k]$.
    Consider the first constraint, which holds over $x\in \C_k\cap\dots\cap \partial \C_1$.
    If $x\in \partial \C_1$, then $\psi_0(x) =0$. 
    If $x\in \C_2$, then $\psi_1(x) \geq 0$, where $\psi_1(x) = \dot{\psi}_0(x) + \alpha_1(\psi_0(x))$. 
    Plugging in~$\psi_0(x) = 0$, we find that
    if $x\in \C_2\cap \partial \C_1$, then $\dot{\psi}_0(x) \geq 0$.
    Then, the $i=1$ constraint is automatically satisfied.

    Consider the $(k-1)^{\text{th}}$ constraint.
    Suppose that the $(k-2)^{\text{th}}$ constraint is satisfied, that is $\dot{\psi}_{k-3}(x) \geq 0 \; \text{for all }x\in \C_k \cap \C_{k-1}\cap \partial \C_{k-2}\cap \dots \cap \C_1.$
    If $x\in \partial \C_{k-1}$, then $\psi_{k-2}(x) =0$.
    If $x\in \C_{k}$, then $\psi_{k-1}(x) \geq 0$, where $\psi_{k-1}(x) = \dot{\psi}_{k-2}(x) + \alpha_{k-1}(\psi_{k-2}(x))$.
    By plugging in~$\psi_{k-2}(x) = 0$, we find that
    $x\in \C_{k}\cap \partial \C_{k-1}$ implies $\dot{\psi}_{k-2}(x)\geq0$.
    Thus, the $(k-1)^{\text{th}}$ constraint is automatically satisfied. 

    For the $k^\text{th}$ constraint, if $x\in\partial \C_k$ then $\psi_{k-1}(x) = 0$.
    However, we do not know if $\dot{\psi}_{k-1}(x) \geq 0$ based on the 
    condition $x\in \partial \C_k\cap\dots\cap \C_1$, and
    the constraint 
    $\dot{\psi}_{k-1}(x) \geq 0$ 
    is not automatically satisfied.
    Thus, the tangent cone condition for the forward invariance of $\cap_{i=1}^k\C_i$ is equivalently expressed as $\dot{\psi}_{k-1}(x)\geq 0$ for all $x\in \partial \C_k$,
    where we use the fact that $\partial\C_k=\partial \C_k\cap \dots\cap\C_1$, because $\C_k\subseteq\dots\subseteq \C_1$.
\end{proof}

The next lemma is used to prove Theorem~\ref{thrm:altHOCBFdef}.

\begin{lemma}\label{lem:inproof}
        If conditions \eqref{eq:thrmpsi0}-\eqref{eq:thrmbeta1par} are satisfied for some $i \in [r] \backslash \{1\}$, then there exists a class $\K$ function $\alpha_{i-1}\in\R[x]$ that (i) is upper-bounded 
        via $\alpha_{i-1}(\psi_{i-2}(x))\leq \big(2\beta_{i-1}(\psi_{i-2}(x))\big)^{1/2}$
        for all $x\in\C_{i-1}$ and
        (ii) 
        satisfies the $i^{\text{th}}$ tangent cone condition from Corollary \ref{corr:tangentCone}, namely, $\dot{\psi}_{i-1}(x)\geq 0 \; \text{for all }x\in\partial \C_i.$
        Therefore, the set $\cap_{k=1}^i\C_k$ is forward invariant for the system.
    \end{lemma}
    
    \begin{proof}
        By Corollary \ref{corr:tangentCone}, the set $\cap_{k=1}^i\C_k$ is forward invariant for the system if and only if
        $\dot{\psi}_{i-1}(x)= \ddot{\psi}_{i-2}(x)+\alpha_{i-1}'(\psi_{i-2}(x))\dot\psi_{i-2}(x)\geq 0\; \text{for all }x\in\partial \C_i.$
        If $x\in\partial \C_i$, then $\psi_{i-1}(x)=\dot\psi_{i-2}+\alpha_{i-1}(\psi_{i-2}(x))=0$. 
        Therefore, we have $\dot{\psi}_{i-2}(x)=-\alpha_{i-1}(\psi_{i-2}(x))$ for all $x\in \partial \C_i$.
        Substituting this into the expression for $\dot\psi_{i-1}$ yields $\dot\psi_{i-1}(x)=\ddot{\psi}_{i-2}(x) -
        \alpha_{i-1}'(\psi_{i-2}(x))\alpha_{i-1}(\psi_{i-2}(x))$
        for all $x\in\partial \C_i$.
        Then the forward invariance condition can be restated as requiring $\ddot{\psi}_{i-2}(x) \geq \alpha_{i-1}'(\psi_{i-2}(x))\alpha_{i-1}(\psi_{i-2}(x))\; \text{for all }x\in \partial \C_i.$
        We can formulate a stronger sufficient condition by requiring
        it to hold over all $x \in \C_{i-1}$ because $\C_i\subseteq \C_{i-1}$.
        Therefore, if there exists a class $\K$ function $\alpha_{i-1}\in\R[x]$ such that 
        \begin{equation}\label{eq:proof4}
            \ddot{\psi}_{i-2}(x) \geq \alpha_{i-1}'(\psi_{i-2}(x))\alpha_{i-1}(\psi_{i-2}(x))\; 
        \end{equation}
        for all $x\!\in\!\C_{i-1}$,
        then $\cap_{k=1}^i\C_k$ is forward invariant for \eqref{eq:dynamics}.

        Let $\bar{\zeta}_{i-1}=\max_{x \in \C_{i-1}}\psi_{i-2}(x)$ and $\beta_{i-1}\in\R[\zeta]$, where $\beta_{i-1}'\in\R[\zeta]$ is the derivative of~$\beta_{i-1}$. 
        If \eqref{eq:thrmbeta1int} holds for $i$, then $\beta_{i-1}'(\zeta)>0$ for all $\zeta\in(0,\bar{\zeta}_{i-1})$. 
        If \eqref{eq:thrmbeta1par} holds for $i$, then $\beta_{i-1}'(0)\geq 0$.
        Therefore, there exists a class $\K$ function $\alpha_{i-1}$ such that (i) $\beta'_{i-1}(\zeta)\geq \alpha_{i-1}'(\zeta)\alpha_{i-1}(\zeta)>0$ for all $\zeta \in(0,\bar{\zeta}_{i-1})$, and (ii) $\beta'_{i-1}(0)\geq \alpha'_{i-1}(0)\alpha_{i-1}(0)=0$ by Lemma \ref{lem:classKprime}.
        If condition \eqref{eq:thrmpsi0} holds for $i$, then $\ddot{\psi}_{i-2}(x) \geq \alpha'_{i-1}(\psi_{i-2}(x))\alpha_{i-1}(\psi_{i-2}(x))$
        for all $x\in \C_{i-1}$. 
        Therefore, the function $\alpha_{i-1}$ satisfies
        the $i^{th}$ tangent cone condition \eqref{eq:proof4}, rendering $\cap_{k=1}^i\C_k$ forward invariant for the system.
        Consider the bound $\alpha'_{i-1}(\zeta)\alpha_{i-1}(\zeta)\leq \beta_{i-1}'(\zeta)$ for all $\zeta \in(0,\bar{\zeta}_{i-1})$.     
        We obtain an upper-bound on $\alpha_{i-1}$ by integrating both sides of this inequality $\int_0^{\zeta}\alpha'_{i-1}(\xi)\alpha_{i-1}(\xi) d\xi \leq \int_0^{\zeta}\beta'_{i-1}(\xi)d\xi,$ 
        resulting in $\frac{1}{2}\alpha^2_{i-1}(\zeta )- \frac{1}{2}\alpha_{i-1}^2(0)\leq \beta_{i-1}(\zeta) -\beta_{i-1}(0).$
        By definition, $\alpha_{i-1}(0)=\beta_{i-1}(0)=0$. Then $\alpha_{i-1}$ is upper-bounded as
        $\alpha_{i-1}(\zeta)\leq \big(2\beta_{i-1}(\zeta)\big)^{1/2}$
        for all $\zeta\in [0,\bar{\zeta}_{i-1}]$.
        Then         
        $\alpha_{i-1}(\psi_{i-2}(x)) \!\leq\! \sqrt{2\beta_{i-1}(\psi_{i-2}(x))} \; \text{for all }x\!\in \!\C_{i-1}.$
    \end{proof}

\subsection{Proof of Theorem~\ref{thrm:altHOCBFdef}} \label{app:altHOCBFdef}
 By Lemma \ref{lem:inproof}, if~\eqref{eq:thrmpsi0}-\eqref{eq:thrmbeta1par} are satisfied for $i=2$, then there exists a class $\K$ function $\alpha_1\in\R[x]$ that is upper-bounded via $\alpha_{1}(\psi_0(x))\leq \sqrt{2\beta_1(\psi_0(x))}\; \text{for all } x\in \C_1$ 
    and the set $\C_2\cap \C_1$ is forward invariant for the system.
    Similarly, if \eqref{eq:thrmpsi0}-\eqref{eq:thrmbeta1par} are satisfied for $i=r$, then there exists a class $\K$ function $\alpha_{r-1}\in\R[x]$ that is upper-bounded via $\alpha_{r-1}(\psi_{r-2}(x))\leq \sqrt{2 \beta_{r-1}(\psi_{r-2}(x))} \; \text{for all }x\in \C_{r-1}$
    and the set $\C_r\cap\dots\cap\C_1$ is forward invariant for the system. 
    This holds for all $i\in[r]\backslash\{1\}$.
    The function $\alpha_r$ appears in the admissible set $\Psi_r(x)=\{u\in\U\mid \dot{\psi}_{r-1}(x,u)+\alpha_r(\psi_{r-1}(x)) \geq 0\}$. 
    If \eqref{eq:thrmpsir-1} holds, then $\dot{\psi}_{r-1}(x,u)\geq -\alpha_r(\psi_{r-1}(x))$ for all $x\in\C_r$. 
    If~\eqref{eq:thrmbetar-1} and \eqref{eq:thrmbetar-1par} hold, then $\alpha_r$ is a class $\K$ function by \cite[Definition 4.1]{khalil2015nonlinear}.
    Therefore, there exists an $\alpha_r$ such that $u(t)\in \Psi_r(x(t)) = \{u\in\U\mid \dot{\psi}_{r-1}(x,u)\geq -\alpha_r(\psi_{r-1}(x))\}$ for all $t\geq 0$, where $u(t)\in\U$ by hypothesis.
    \hfill $\blacksquare$

\subsection{Proof of Theorem \ref{thrm:singleSynth}}
\label{app:singleSynth}

The first $r-1$ SOSPs in part (a) validate $\psi_0$ using Theorem \ref{thrm:altHOCBFdef}.
    Let $S_{i-1} = \{\psi_{i-2},h_1,\dots,h_p\}$, where the semialgebraic set generated by $S_{i-1}$ is $K_{S_{i-1}}=\{x\in\R^n : \psi_{i-2}(x) \geq 0,h_1(x)\geq0,\dots,h_p(x)\geq0\}$ and the quadratic module generated by $S_{i-1}$ is $Q_{S_{i-1}}=\{q\in\R[x] : q(x) = s^{i-1}_0(x) + s^{i-1}_1(x) \psi_{i-2}(x)+(s^{i-1}_2(x))^Th(x)\}$, with $s^{i-1}_0,s^{i-1}_1\in\Sigma[x]$ and $s^{i-1}_2\in\Sigma^p[x]$. 
    Consider $\ddot{\psi}_{i-2}-\beta_{i-1}'(\psi_{i-2})\in Q_{S_{i-1}}$, i.e., \eqref{eq:SSi} is satisfied
    and~$\ddot{\psi}_{i-2}-\beta'_{i-1}(\psi_{i-2})$ can be written in the form $\ddot{\psi}_{i-2}(x) - \beta_{i-1}'(\psi_{i-2}(x)) = s^{i-1}_0(x) + s^{i-1}_1(x)\psi_{i-2}(x) + (s^{i-1}_2(x))^Th(x)$.
    Then the function $\ddot{\psi}_{i-2}(x)\geq \beta_{i-1}'(\psi_{i-2}(x))$ for all $x\in K_{S_{i-1}}$ and all $i\in[r]\backslash\{1\}$, where  $K_{S_{i-1}}=\C_{i-1}\cap \X$.
    By construction of $\beta_{i-1}'$, 
    the inequalities $\beta_{i-1}'(\psi_{i-2}(x)) > 0 \quad \text{for all }x\in \text{int}(\C_{{i-1}})$ and $\beta_{i-1}'(\psi_{i-2}(x))\geq 0 \quad \text{for all }x\in \partial \C_{i-1}$
    hold for all $i\in[r]\backslash\{1\}$.
    Therefore, if the SOSP in part (a) is feasible for all $i\in[r]\backslash\{1\}$, 
    then \eqref{eq:thrmpsi0}-\eqref{eq:thrmbeta1par} in Theorem \ref{thrm:altHOCBFdef} are satisfied.
    This implies that there exist class $\K$ functions $\alpha^\star_{i-1}\in\R[x]$ for $i\in[r]\backslash\{1\}$ such that $\alpha^\star_{i-1}(\psi_{i-2}(x)) \leq \big(2\beta^\star_{i-1}(\psi_{i-2}(x))\big)^{1/2}$
    for all $x\in \C_{i-1}\cap \X$.
    Then by Lemma \ref{lem:inproof} the set $\cap_{i=1}^r\C_i$ is forward invariant for the system.

    The SOSP in part (b) is the final step in validating $\psi_0$ with Lemma \ref{lem:hocbf}.
    Let $S_r=\{\psi_{r-1},h_1,\dots,h_p\}$, where the semialgebraic set generated by $S_r$ is $K_{S_r}=\{x\in\R^n : \psi_{r-1}(x)\geq0,h_1(x)\geq0,\dots,h_p(x)\geq0\}$ and the quadratic module generated by $S_r$ is $Q_{S_r}=\{q\in\R[x] : q(x) = s^r_0(x) + s^r_1(x) \psi_{r-1}(x) + (s^r_2(x))^Th(x)\}$, with $s^r_0,s^r_1\in\Sigma[x]$ and $s^r_2\in\Sigma^p[x]$.
    Consider $\dot{\psi}_{r-1}+\alpha_r(\psi_{r-1}(x))\in Q_{S_r}$, i.e., \eqref{eq:SSr} is satisfied and $\dot{\psi}_{r-1}+\alpha_r(\psi_{r-1}(x))$ can be written in the form $\dot{\psi}_{r-1}(x,u(x))+\alpha_r(\psi_{r-1}(x)) = s^r_0(x) + s^r_1(x)\psi_{r-1}(x) +(s^r_2(x))^Th(x).$
    Then $\dot{\psi}_{r-1}(x,u(x)) \geq -\alpha_r(\psi_{r-1}(x))$ for all $x\in K_{S_r}=\C_r\cap \X$.
    By construction of $\alpha_r$, both \eqref{eq:thrmbetar-1} and \eqref{eq:thrmbetar-1par} are satisfied. 
    Then there exists a class $\K$ function $\alpha_r\in\R[x]$ such that $u^\star(x)\in \{u\in\R^m : \dot{\psi}_{r-1}(x,u)\geq -\alpha_r(\psi_{r-1}(x))\}$ for all $x\in\C_r\cap \X$.

    Let $Q_u$ be the $q-$dimensional quadratic module generated by $S_r$, namely $Q_u=\{q_u\in\R^q[x] : q_u(x) = \tilde{s}_0(x) + \tilde{s}_1(x)\psi_{r-1}(x) + (\tilde{s}_2(x))^Th(x) \}$, where $\tilde{s}_0,\tilde{s}_1\in\Sigma^q[x]$ and $\tilde{s}_2\in\Sigma^{p\times q}[x]$. 
    Consider $c(u(x))\in Q_u$, i.e., \eqref{eq:SSU} is satisfied and $ c(u(x)) = \tilde{s}_0(x) + \tilde{s}_1(x)\psi_{r-1}(x) + (\tilde{s}_2(x))^Th(x)$.
    Then $c(u(x))\geq 0$ for all $x\in \C_r\cap \X$ by Lemma \ref{lem:nonneg}.
    Therefore, $u^\star(x)\in\U$ for all $x\in \C_r\cap \X$.
    Combined with the fact that $u^\star(x)\in\{u\in\R^m : \dot{\psi}_{r-1}(x,u)\geq -\alpha_r(\psi_{r-1}(x))\}$ for all $x\in \C_r\cap\X$, Theorem \ref{thrm:altHOCBFdef} is satisfied. Therefore, $u^\star(x)\in \Psi_r(x)=\{u \in \U : \psi_r(x,u)\geq 0\}$ for all $x\in \C_r\cap \X$.
    
    In part (b), the constraint \eqref{eq:SSV} and the objective function \eqref{eq:SSSobjfun2} determine the measure of safe stabilization $\rho^\star$ that the system can maintain.
    Let $S_r$, $K_{S_r}$, and $Q_{S_r}$ be as defined above.  
    If $-L_fV-L_gVu+\rho\in Q_{S_r}$, i.e., if \eqref{eq:SSV} is satisfied and we can write $-L_fV(x) - L_gV(x)u(x) + \rho = \hat{s}_0(x) + \hat{s}_1(x) \psi_{r-1}(x) +(\hat{s}_2(x))^Th(x),$
    then $L_fV(x) + L_gV(x) u(x) \leq \rho$ for all $x\in \C_r\cap \X$.
    
    In total, the feasibility of the sequence of SOSPs in the theorem statement provides a guarantee of continued feasibility for the SQP in \eqref{eq:SQP} in the sense of Definition \ref{def:guarantee},  an upper bound for the class $\K$ functions $\alpha^\star_i$ as a function of $\beta^\star_i$ for $i\in[r-1]$, and the class $\K$ function $\alpha_r^\star$. \hfill $\blacksquare$

\subsection{Proof of Lemma \ref{lem:multiHOCBF}}
\label{app:multiHOCBF}
By Lemma \ref{lem:Nagumos}, the set $C_\sys=C^1\cap\dots\cap C^J$ is forward invariant for system \eqref{eq:dynamics} if and only if~$\dot{x}\in T_{C_\sys}(x)$, i.e., $\dot{x} \in \{z\in\R^n \mid \nabla\psi_{r_j-1}^j(x)^Tz\geq 0 \; \text{for all }j\in \text{Act}(x) \}$
    for all $x\in \partial C_\sys=\partial (\C^1_{r_1}\cap\dots\cap\C^J_{r_J})$,
    where $\text{Act}(x)= \{j\in[J] \mid \psi^j_{r_j-1}(x)=0\}$.
    The boundary of the safe set is equal to 
    $\partial C_\sys = \partial (\C^1_{r_1}\cap\dots\cap\C^J_{r_J}) = (\partial \C^1_{r_1}\cap C_\sys) \cup \dots \cup (\partial \C^J_{r_J}\cap C_\sys).$
    Let~$u$ be a Lipschitz continuous controller
    with~$u(x) \in \Psi_{\bar{r}}(x)$ for all~$t$. 
    Under~$u$, for each~$j \in [J]$ we have~$\psi^j_{r_j}(x,u) \geq 0$.
    Expanding gives $\dot{\psi}^j_{r_j-1}(x,u) + \alpha^j_{r_j}\big(\psi^j_{r_j-1}(x)\big) \geq 0. $
    For any $x \in \partial \C^j_{r_j}$, we have $\psi^j_{r_j-1}(x) =0$ and
    thus $\dot{\psi}_{r_j-1}^j(x,u) \geq 0$ for all 
        $x \in \partial \C^j_{r_j} \cap C_\sys$
    and each $j\in[J]$.
    The result follows 
    by using the expanded form of~$\partial C_{\sys}$. \hfill $\blacksquare$

\end{document}